# Multi-dimensional laser mode combs (mode hyper-combs)

Alon Schwartz and Baruch Fischer

Department of Electrical Engineering, Technion, Haifa 32000, Israel

E-mail: fischer@ee.technion.ac.il

**Abstract:** Laser frequency combs, as most lasers, are one-dimensional. Here we present a realization of d-dimensional laser mode lattices (mode hyper-combs) with unique properties. They are constructed from regular 1-dimensional combs by multi-frequency modulation in active mode-locking (AML). The hyper-comb, with near neighbor mode interaction and noise functioning as temperature, is mapped to interacting magnetic spin-lattices in the spherical-model, which is one of the few statistical-mechanics systems soluble in all dimensions. The important result is that such systems have, in $d>2$ dimensions, a phase-transition to a global mode-phase-ordered hyper-comb. It changes the nature of AML lasers, giving ultimately short and robust pulses which can capture very broad frequency bandwidths. Additionally, the hyper-combs can serve as a rare physical realization of the spherical-model in any dimension.

**Introduction:** Ultra-short laser pulses that reach the few femtosecond regime [1,2] are a key platform for advancing important fields like metrology [3-5] and attosecond science [5-9]. They are based on broad frequency combs [3,4] of phase-aligned axial-modes, usually obtained by mode-locked lasers [1,2,10]. Those combs (as mostly lasers) are one-dimensional.

Broad frequency combs are mostly obtained by passive mode-locking (PML) [1-10] which produces ultra-short pulses with durations that can reach a few optical cycles. Active mode-locking (AML) [10,11] is another way to achieve short pulses, generally with longer durations compared to PML. There is an inherent difference between the two methods, explainable by statistical light-mode dynamics (SLD) [12-20]. SLD treats the laser as a many-body system in a statistical-mechanics approach, where the modes are the "particles" and noise has the role of temperature. The mode phases replace the spin orientations of the magnetic spin case. On that basis, the pulsation in PML was shown [12-15] to be a first-order phase-transition from phase-disordered to ordered mode system. PML lasers were also shown to exhibit critical behavior [16,17]. On the other hand, in AML under regular modulation, no global mode ordering exists at any finite noise level [18]. It falls in the category of 1-dimensional many-body systems with short-range interaction, that do not have phase-transitions and global ordering at any temperatures, but zero, due to coupling (energy) versus entropy considerations. It means that weak noise ("temperature"), even spontaneous emission, can affect the phase alignment of a long but fragile mode chain (comb), preventing a global mode-phase ordering [18], especially for very broad frequency bandwidths. PML is different due to the saturable-absorber that causes an effective long-range interaction between all modes [12-15]. Also complex AML modulations can result condensation-like behavior [19,20]. With regular modulations, the 1-dimensional laser frequency-comb produced by AML was mapped [18] to the 1-dimensional spherical model [21-25], a variant of the Ising model for magnetic spins with nearest-neighbor interaction that was studied in statistical mechanics and was solved analytically in all dimensions. Similarly to the spherical model, there is no constraint in the laser system on the power of each individual mode, but there is one on the overall power of all modes.

Since the dimensionality has a major importance in statistical-mechanics system, the comb fragility of AML in 1-dimension may be eliminated at higher dimensions. Indeed, in the spherical model, in a dimension higher than 2 (d > 2, excluding 2; d in statistical-mechanics calculations is not restricted to integers) there is a second-order phase-transition from disordered spin phase to long-range ordering, due to the high-dimensional connectivity. A direct way to obtain higher than 1-dimensional comb in lasers would be to use higher dimensional cavities which are not easily achievable with mode-coupling. The idea in this work is to construct a d-dimensional mode hyper-comb with mode-interaction from 1-dimension. We proceed with the way to do that.

**The mode hyper-comb construction:** The d-dimensional interacting mode-combs are obtained by a remarkably simple way: AML with a multi-frequency modulation, $d$ frequencies for $d$ dimensions. The first frequency matches the basic cavity resonance $\Omega_0 = c / nl$ that usually falls in the RF regime, ($l$ is the cavity roundtrip length and $n$ is the refractive index), We then add more terms with multiples of the lower frequencies, having altogether: $m_1\Omega_0$, $m_2\Omega_0$,... $m_d\Omega_0$ (here $m_1 = 1$). Each modulation frequency $\Omega_n = m_n\Omega_0$ ($n$=1,2,…) induces coupling between modes $i$ and $i \pm m_n$. (We assume cosinusoidal amplitude modulation that produces two sidebands around each mode, but any higher, but finite number of sidebands will not change the basic results.) For *d=1* (with $m_1 = 1$), we have the regular 1-dimensional comb with induced nearest-neighbor coupling between modes $i$ and $i \pm 1$. A second modulation frequency $\Omega_2 = m_2\Omega_0$ induces additional interactions between modes $j$ and $j \pm m_2$. Therefore, we divide the 1-dimensional comb into segments, each with $m_2$ modes, and fold them to obtain additional rows and a 2-dimensional lattice mode structure $a_{i,j}$, as shown in Figs. 1a and 1b. $a_{i,j}$ denotes the complex amplitude of the laser light modes. In that lattice, each mode $a_{i,j}$ is coupled to its four nearest-neighbor modes $a_{i\pm1, j\pm1}$. With a third modulation having a higher harmonic frequency which is also a multiple of the second frequency, we apply a similar procedure, illustrated in Fig. 1c. Now, successive folding of two-dimensional arrays of ($m_2 \times \frac{m_3}{m_2}$) modes leads to a 3-dimensional lattice of modes $a_{i,j,k}$, each coupled to its six nearest-neighbor modes $a_{i\pm1, j\pm1, k\pm1}$. This restructuring procedure can proceed to higher dimensions, achievable by including in the modulation more frequency terms. We stress that it is not just an arbitrary or formal restructuring. We obtain a d-dimensional mode lattice (hyper-comb) with nearest-neighbor coupling just as we have in solid state atomic or spin lattices. It is also possible to construct different kinds of hyper-combs, such as the triangular (hexagonal) mode-lattice shown in Fig. 2, by using three modulating frequencies. Then, $m_{i+1}/m_i$ are not necessarily integers. In Fig. 2, $m_3/m_2 = m_3/(m_3 + 1) = 8/7$.

We note that the hyper-comb is reconstructed from 1-dimension, and is not a regular mode structure in $\bar{\mathbf{k}}$ space. Real $\bar{\mathbf{k}}$ space mode-combs can be obtained in two or three-dimensional laser cavities but are difficult to obtain experimentally with mode-coupling. In our conceptual lattice the inter-mode distance, measured in frequency terms, is given by:

$$\nu_{ijk...} - \nu_{i'j'k'...} = [(i-i')m_1 + (j-j')m_2 + (k-k')m_3 +...]\Omega_0, \qquad (1)$$

($m_1 = 1$ when the first frequency modulation is taken to be the cavity resonance.) Thus, the frequency difference between nearest-neighbor modes along the $i+1$ axis is higher by a factor $N_i = m_{i+1}/m_i$ from the nearest-neighbors difference along $i$.

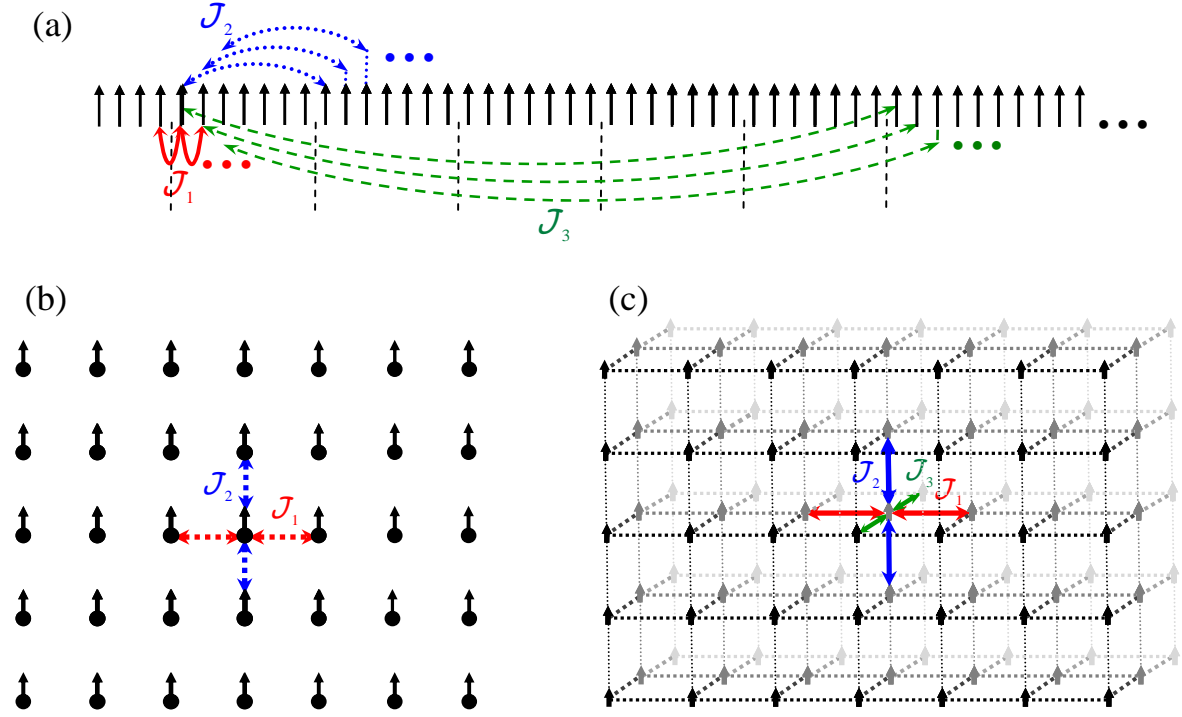

**Fig. 1. Construction of the d-dimensional mode hyper-comb:**

(The small arrows at the lattice points denote the modes phasors, where the directions indicate the phases. They are all drawn in the same direction, in an ordered phase, but they can be disordered, as discussed in the paper. The comb size in the figure serves only for illustration.)

(a) 1-dimensionsl mode-comb: AML modulation with three frequencies, $\Omega_n = m_n \Omega_0$ (in the figure $m_1 = 1$, $m_2 = 7$, $m_3 = 35$), induces coupling between modes $j$ and $j \pm m_n$. The corresponding coupling strengths are $\mathcal{J}_1$, $\mathcal{J}_2$ and $\mathcal{J}_3$.

(b) 2-dimensionsl mode-comb: The second modulation frequency $\Omega_2 = m_2 \Omega_0$ ($m_2$=7 in the figure) induces coupling between each mode $j$, in a segment of $m_2$ modes, and mode $j \pm m_2$ in the following segment of $m_2$ modes. Upon folding and placing the second segment as a second row above the first one, the coupled modes become nearest neighbors. Repeating the folding procedure for the next segments gives a two-dimensional comb where each mode (except those on the boundary) is coupled to its four nearest neighbor modes.

(c) 3-dimensionsl mode-comb: A third modulation frequency $\Omega_3 = m_3 \Omega_0$ ($m_3$=35 in the figure, has to be a multiple of $m_2$) couples mode $j$ and mode $j \pm m_3$. Then the folding is of 2-dimensional arrays (each of $7 \times \frac{35}{7} = 7 \times 5$ modes in the figure). It gives a 3-dimensional comb, where each mode is coupled to its six nearest neighbor modes.

This procedure can be proceeded to higher dimensions, depending on the available frequency bandwidth and the number of modes taken for each dimension.

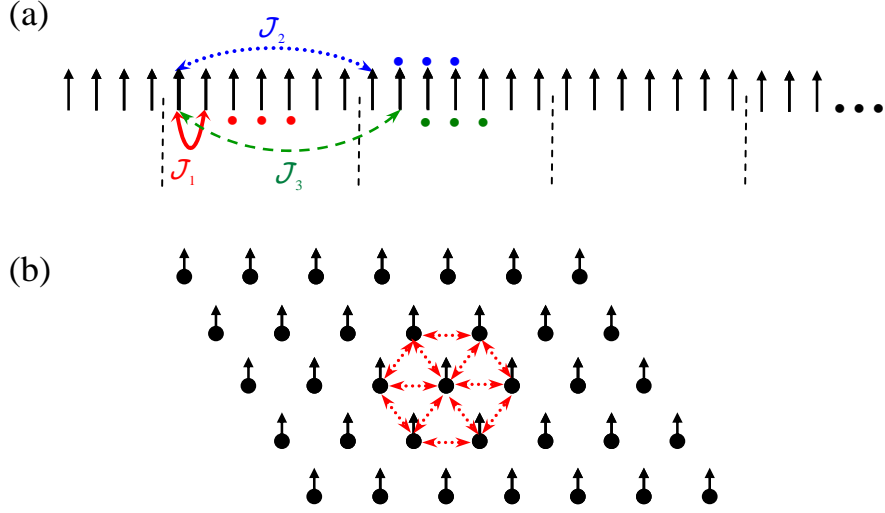

**Fig. 2. Triangular (hexagonal) mode-lattice:**

(a) 1-dimensionsl mode-comb with three frequencies, $\Omega_1 = \Omega_0$, $\Omega_2 = m_2\Omega_0$, and $\Omega_3 = m_3\Omega_0 = (m_2+1)\Omega_0$. (in the figure $m_1 = 1$, $m_2 = 7$, $m_3 = 8$).

(b) Reconstruction to a 2-dimensionsl triangular (hexagonal) mode-lattice. In the reconstruction procedure each additional segment that is placed as a new row is horizontally shifted by half a period. It reflects the coupling of each mode in the former row to two adjacent modes in the above row, besides the coupling to nearest neighbor modes in the same row. Experimentally, those couplings can be different in strength and sign.

**The mode hyper-comb and the spherical-model**: The d-dimensional spherical model is one of the few (if not the only) model that have analytical solutions in all dimensions [21-25]. It exhibits for d>2 a second-order phase-transition from disordered to ordered spin phase at a finite temperature. Since the laser mode system is mapped to the spherical spin model, we can say that AML at d>2 has a phase-transition to a phase-ordered mode structure. The high-dimensional mode connectivity overcomes the entropy randomizing force in the free energy, yielding a transition to an ordered mode-phase structure with enhanced resistance against noise, just as it happens in thermodynamics for interacting particles in solids and magnetic spin systems. In the time domain, the robust phase ordered mode-comb then gives ultimately short pulses that can use very broad frequency bandwidths.

The AML equation of motion for the complex amplitudes $a_{ijk...}$ with noise is given by [18]:

$$\frac{\partial a_{ijk...}}{\partial \tau} = \sum_{n=1}^{d} \frac{A_n}{2} \sum_{i,j,k} a_{i\pm1, j\pm1, k\pm1...} + [g(P)-l]a_{ijk...} + \Gamma_{ijk...}(t,\tau), \qquad (2)$$

where τ is the long term time variable that counts the round-trips of light in the laser cavity, $A_m$ is the complex modulation amplitude at frequency $\Omega_n$, $g(P)$ is the total power $P$ dependent slow (compared to one roundtrip) saturable gain, and $l$ is the loss in the cavity. The equation does not include dispersion. $\Gamma_{ijk...}$ is an additive noise term that can originate from spontaneous emission, or any other internal or external source, modeled by a white Gaussian noise process with covariance $2T$: $<\Gamma_{ijk...}(\tau)\Gamma^*_{ijk...}(\tau')> = 2T\delta(\tau-\tau')\delta_{ii'}\delta_{jj'}\delta_{kk'}...$, $<\Gamma_{ijk...}(\tau)> = 0$, where $<\>$ denotes average, and $T$ has the role of temperature. Eq. 2 can be written as:

$$\partial a_{ijk...} / \partial \tau = -\partial H / \partial a^*_{i\pm1,j\pm1,k\pm1...} + \Gamma(t,\tau)_{ijk...}$$
$$\partial a^*_{ijk...} / \partial \tau = -\partial H / \partial a_{i\pm1,j\pm1,k\pm1...} + \Gamma^*(t,\tau)_{ijk...} \quad , \quad (3)$$

where $H$ is a Hamiltonian-like quantity (here lossy and imaginary[17,23]):

$$H = H_{int} - g_0 \ln(P_{sat} + P) + lP \quad (4)$$

and $H_{int}$ is the mode-interaction part:

$$H_{int} = -\sum_{n=1}^{d} \frac{A_n}{2} \sum_{i,j,k} a_{ijk...} a^*_{i\pm1,j\pm1,k\pm1...} \quad , \quad (5)$$

We note that in Eq. 2 we took a flat profile [18] for the spectral filtering and the gain spectrum that is justified for broad bandwidths and enables the exact mapping to the spherical model. It many cases it is taken to be parabolic, but apart from some effect on the pulse shape, it gives the basic features of the mode system and the pulses. As done in Ref. 18, by deriving the corresponding Fokker-Planck equations, and assuming a stabilized total power $P = \sum_{i,j,k...} a_{ijk...} a^*_{ijk...} \to P_0$, we have for the steady state mode distribution:

$$\rho(a_{ijk...},...a_{ijk...}) \propto \delta(P - P_0) \exp(-H_{int}/2T). \quad (6)$$

Therefore the mode system configuration obeys Gibbs-like statistics [18], with $H_{int}$ and the noise $T$ functioning as the (interaction) energy and temperature. This is the central base that connects the mode system to statistical mechanics [12-18]. The global constraint on the mode power: $\sum_{i,j,k...} a_{ijk...} a^*_{ijk...} = P_0$, makes it similar to magnetic spin system in the spherical model. The mapping is simply done [18] by redefining $\tilde{a}_{ijk...} = (N/P_0)^{1/2} a_{ijk...}$ for the mode complex amplitudes that replace the spins. Then we rewrite Eq. 5: $H_{int} = -\sum_{n=1}^{d} \mathcal{J}_n \sum_{i,j,k} \tilde{a}_{ijk...} \tilde{a}^*_{i\pm1,j\pm1,k\pm1...}$ with a normalized mode-coupling parameter $\mathcal{J}_n = A_n P_0 / 2N$. For simplicity, we assumed that all modulation frequencies have the same phases and amplitudes ($A_n = A$). Then the coupling (generally complex) in all axes are equal, $\mathcal{J}_n = \mathcal{J}$ (isotropic crystal), but it can be generalized to anisotropic coupling. $\mathcal{J}_n$ can be easily varied and also have different relative signs (like in anti-ferromagnetic spins) in one experiment by changing the corresponding amplitudes and phases. We also note that in experiments with lasers, it is the power $P_0$ that is usually varied, although $T$ can also be changed by noise injection[18-19] as well as the modulation amplitude $A$.

According to the spherical model [21-25], the mode system for $d > 2$ undergoes a second-order phase-transition. It formally occurs in the thermodynamic limit, but practically finite hyper-comb sizes can be sufficient (discussed in methods). The average (over magnitude and phase) normalized mode complex amplitude $\tilde{a}_{ijk}$ is given by [23,24]:

$$<\tilde{a}_{ijk}> = \begin{cases} (1 - K_c/K)^{1/2} & for\ K \geq K_c \\ 0 & for\ K < K_c \end{cases}. \quad (7)$$

The parameters and quantities are discussed in the methods section: $K = \mathcal{J}/T$, $\mathcal{J} = AP_0/2N$, is the mode nearest-neighbor coupling strength, $P_0$ - the laser total power, $A$ - the modulation amplitude, $N$ - the overall mode number, and $T$ is the noise strength ("temperature"). The second order transition occurs at $K_c = (\mathcal{J}/T)_c = b_d$, where $b_d = \int_0^\infty \exp(-dx)[J_0(ix)]^d \, dx$, for $d>2$, ($b_3 \approx 0.505$, $b_4 \approx 0.310$, $b_5 \approx 0.231$), and $J_0$ is the zero order Bessel function. Eq. 7 describes a second-order phase transition at $K = K_c$, corresponding to a critical exponent [24] $\beta = 1/2$. Since in lasers we usually vary the laser power $P_0 = \sum_{i,j,k...} a_{ijk...} a^*_{ijk...}$, it is useful to write Eq. 7 explicitly with $P_0$:

$$<a_{ijk}> = \begin{cases} (P_0/N)^{1/2}(1-P_c/P_0)^{1/2} & \text{for } P_0 \geq P_c \\ 0 & \text{for } P_0 < P_c \end{cases} \quad (8)$$

Therefore, above $K_c$ (above $P_c$ when $T$ is kept constant or below $T_c$ when $P_0$ is constant), the mode system becomes ordered with long-range mode-mode correlation that was shown [21] to be for $d=3$: $<\tilde{a}_{ijk}\tilde{a}^*_{i'j'k'}> = (1-K_c/K) + C[(i-i'),(j-j'),(k-k')]$. It consists of a mean value, independent on mode distance, plus a function C that decays with the distance due to the fluctuations. It means that for $K \geq K_c$, the constant correlation, that increases to 1 as $K$ grows, is between all modes, so that $N_{cor} = N$.

Below $K_c$ for $d > 2$ or any $K$ for $d \leq 2$, $<\tilde{a}_{ijk}> = 0$, but the mode correlation is nonzero but decays with increasing mode distance. For example, in 1-dimension the correlation decays exponentially and the correlation length, in terms of modes, was shown to be [18] $N_{cor} \approx 2K$. In the mode hyper-comb, for $d=3$ below the phase-transition, the correlation is again decaying to zero with increasing distance, and the correlation length is [24] $N_{cor} \propto (1-K/K_c)^{-1}$, (it corresponds to critical exponent $\nu = 1$). Therefore as $K \to K_c$, $N_{cor}$ diverges [24], and practically reaches the maximum possible value $N_{cor} \to N$, but at the same time there is a significant increase of fluctuations [24] (most meaningfully of the modes phases), evidenced by having here $<\tilde{a}_{ijk}> = 0$.

**Pulse generation of the mode hyper-comb:** Experimentally, the correlation and mode-ordering in the comb is measured in the time domain. In atomic and spin systems the parallel measurement is of electromagnetic-wave or neutron scattering [24]. We saw that for $K \geq K_c$ the correlation extends to the whole system and its magnitude increases with $K$. Therefore, the pulse captures the full frequency bandwidth of all $N$ modes, resulting in an ultimate pulse-width $\tau_{pulse} \approx 2\pi/(\Omega_0 N)$. However, even where $<\tilde{a}_{ijk}> = 0$ the correlation makes the laser generate pulses. In 1-dimensional comb it is limited to $\tau_{pulse} \approx 2\pi/(\Omega_0 N_{cor}) \approx \pi/(\Omega_0 K)$. (The pulse shape was shown to be there Lorenzian [18], similarly to the scattering angle experiments for particles and spins [24], resulting from the assumption of flat spectral filtering and gain profiles.) For d>2, even below the phase-transition, as $K \to K_c$ the correlation length diverges, meaning that already at $K_c$ it extends to the whole system. Nevertheless, at this stage the mode system is accompanied by strong fluctuations [24] that affect the pulses. Above the phase-transition, the fluctuations are depressed as $K$ increases beyond $K_c$, and therefore the pulses can be optimal. This is the important result obtained by the multi-dimensional AML mode hyper-comb. The phase-transition presence occurring at d>2 makes the system fall into an ordered phase, regardless the noise, and therefore produce pulses that use the full frequency bandwidth. It is in contrast to 1-dimensional combs that are fragile due to even weak noise. Their correlation length still increases with *K*, but in a gradual way.

In the mode hyper-comb we have to be careful how to transform the inter-mode distances to frequencies (given in the methods section), and then to the time domain, since the above correlation lengths are given in terms of number of modes. For the pulse-width we take the largest correlation length in terms of frequency bandwidth, which is along the d axis (*i=d*, in the folding procedure of the hyper-comb construction).

The pulse rate issue is also interesting. Despite taking for the pulse-width the d axis mode separation, which is the highest difference in terms of frequency, the closest near-neighbor modes (along the x axis) remain coupled and dominate the pulse rate. Therefore, there is a basic difference between the hyper-comb and 1-dimensional AML with only one high harmonic modulation, say the highest one in the hyper-comb ($\Omega_d = m_d \Omega_0$). In a hyper-comb at $K > K_c$, all modes along all axes are coupled and ordered, and the full bandwidth generates pulses at the fundamental rate $\Omega_1 = \Omega_0$ of a single oscillating pulse in the cavity. Another way to realize that is in the time domain. Pulses are generated near the global maximum of the overall modulation waveform (the sum of all frequencies). It usually consists of one or two closely spaced maxima, at or near the maximum of the basic modulation frequency. Therefore, we will have one pulse oscillation ("singlet" or "doublet") at the basic frequency rate $\Omega_1 = \Omega_0$. In a regular single high ($m_d$) harmonic frequency modulation, however, we have multi-pulse oscillation with a rate $\Omega_d = m_d \Omega_0$. Then the coupling is only between modes spaced $m_d$ modes apart, thus having $m_d$ interleaved sub-combs ("supermodes" [26]), usually without phase coupling between them [26], resulting in non-optimized pulses.

We turn to quantity sides. It was noted [18] that for the 1-dimensional AML comb, weak noise, even spontaneous emission, can destabilize the phase alignment of long mode chains, and therefore the mode-correlation length $N_{cor} \approx 2K = 2AP_0/W$ is limited. ($W = 2NT$ is the total noise power in the frequency band). For example, in an erbium-doped fiber laser, with a cavity length of 10m that gives $\Omega_0/2\pi = 2 \cdot 10^7 Hz$ ($n \approx 1.5$), and a modulation power over total noise ratio $K = AP_0/W = 50$ we have $N_{cor} \approx 100$, and $\tau \sim 10^{-9}$ sec. In the hyper-comb for that *K* the laser lies deeply in the ordered phase (for $d = 3$, $K_c \approx 0.505$.) Then the pulse-width is given by the full frequency bandwidth. Taking for it $10^{13} Hz$, gives $N \sim 5 \cdot 10^5$, and $\tau_{pulse} \sim 10^{-13}$ sec, comparable to what is obtained by PML in erbium-doped fiber lasers. Larger bandwidths in various lasers can lead to femtosecond pulses.

**Hyper-comb size:** The mode length at each dimension *i* is given by $N_i = m_{i+1}/m_i$ for $i = 1,...(d-1)$, and $N_d = N/m_d$, where $N = N_1 \cdot N_2 \cdots N_d$ is the total number of modes in the lattice. The formal thermodynamic limit requires $N \to \infty, N_i \to \infty$. In experiments, the hyper-combs will have a finite size (finite $N_i$), depending on the available bandwidths and the possibility to use high harmonic modulation. In various lasers it can be realistic to have 20-100 modes in each dimension, that can already be regarded as a many body system [18,19]. We note that the folding procedure eliminates the bond(s) at the boundary between segments (arrays), but it doesn't have a significant effect on the calculation. We also note that since phase-transition occurs in the spherical model at *d>2*, we can expect that in 2-dimensional mode-combs, only a few additional layers in the third dimension can be sufficient for getting a phase-transition to the phase-ordered comb.

**Conclusion:** we have presented a unique d-dimensional AML mode-comb with nearest-neighbor mode-coupling, mapped to the exactly soluble spherical model of spins in statistical mechanics. It changes the nature of AML due the phase-transition, providing pulses at the basic cavity rate that can optimally use large frequency bandwidths of lasers. It thus solves an inherently weak side of AML compared to PML, while benefitting from its active modulation properties. We finally note that although the spherical model is exactly soluble in all dimensions, it is usually regarded as unphysical because of the unusual constraint on the overall spin value. The AML laser therefore offers a rare experimental realization of the model.

**Acknowledgments:** This research was supported by the US-Israel Binational Science Foundation (BSF). We thank Alexander Bekker, Gilad Oren and Oded Fischer for valuable help.

**References:**

1. T. R. Schibli, J. W. Kuzucu, J.W. Kim, E. P. Ippen, J. G. Fujimoto, F. X. Kaertner, V. Scheuer, and G. Angelow, "Toward Single-Cycle Laser Systems", *IEEE J. Select. Topics in Quantum. Electron.* **9**, 990 (2003).
2. F. X. Kaertner, E. P. Ippen, and S. Cundiff, "Femtosecond Laser Development; Femtosecond Optical Frequency Comb Technology", Chap. 2, Eds. J. Ye and S. Cundiff, Springer Verlag (2005).
3. Th. Udem, R. Holzwarth, and T. W. Hansch, "Optical frequency metrology", *Nature,* **416**, 233 (2002).
4. D. Jones, S. Diddams, J. Ranka, A. Stentz, R. Windeler, J. L. Hall, and S. T. Cundiff, "Carrier envelope phase control of femtosecond mode-locked lasers and direct optical frequency synthesis", *Science*, **288**, 635 (2000).
5. M. Hentschel, R. Kienberger, Ch. Spielmann, G. A. Reider, N. Milosevic, T. Brabec, P. B. Corkum, U. Heinzmann, M. Drescher, F. Krausz, „Attosecond metrology", *Nature* **414**, 509.
6. Z. Chang, and P. B. Corkum, "Attosecond photon sources: the first decade and beyond", *J. Opt. Soc. Am. B*, **27**, B9 (2010).
7. J. Itatani, F. Quéré, G.L. Yudin, M. Yu Ivanov, F. Krausz, and P. B. Corkum, „Attosecond streak camera", *Phys. Rev. Lett.* **88**, 173903 (2002).
8. H. C. Kapteyn, O. Cohen, I. Christov, and M. M. Murnane, "Harnessing Attosecond Science in the Quest for Coherent X-rays", *Science* **317**, 775 (2007).
9. P. Popmintchev, M. C. Chen, P. Arpin, M. M. Murnane, and H. C. Kapteyn, "The Attosecond Nonlinear Optics of Bright Coherent X-Ray Generation", *Nature Photonics*, **4**, 822 (2010).
10. H. A. Haus, "Mode-locking of lasers", *IEEE J. Sel. Top. Quantum Electron.* **6,** 1173 (2000).
11. D. J. Kuizenga, and A. E. Siegman, "Modulator frequency detuning effects in the FM-mode-locked laser", *IEEE J. Quantum Electron.* **QE-6,** 803 (1970).
12. A. Gordon, and B. Fischer, "Phase transition theory of many-mode ordering and pulse formation in lasers", *Phys. Rev. Lett.* **89,** 103901 (2002).
13. A. Gordon, B. Vodonos, V. Smulakovski, and B. Fischer, "Melting and freezing of light pulses and modes in mode-locked lasers", *Opt. Express* Vol. **11,** 3418-3424, 2003;
14. B. Vodonos, R. Weill, A. Gordon, V. Smulakovsky, A. Bekker, O. Gat, and B. Fischer, "Formation and annihilation of laser light pulse quanta in a thermodynamic-like pathway"; *Phys. Rev. Lett.*, Vol. 93, 153901, 2004
15. O. Gat, A. Gordon, and B. Fischer, "Light-mode locking: a new class of solvable statistical physics systems", *New J. Phys* **7**, 151 (2005).
16. R. Weill, A. Rosen, A. Gordon, O. Gat, and B. Fischer, "Critical behavior of light in mode-locked lasers", *Phys. Rev. Lett*. **95**, 013903 (2005).

17. A. Rosen, R. Weill, B. Levit, V. Smulakovsky, A. Bekker, and B. Fischer, "Experimental observation of critical phenomena in a laser light system", *Phys. Rev. Lett*. Vol. 105, 013905 (2010);
18. A. Gordon, and B. Fischer, "Statistical mechanics theory of active mode locking with noise"; *Opt. Lett*. Vol. 29, p. 1022, 2004.
19. R. Weill, B. Fischer, and O. Gat, "Light-mode condensation in actively mode-locked lasers", *Phys. Rev. Lett*. **104**, 173901 (2010);
20. R. Weill, B. Levit, A. Bekker, O. Gat, and B. Fischer, "Laser light condensate: Experimental demonstration of light-mode condensation in actively mode locked laser", *Opt. Express*, **18**, p. 16520 (2010).
21. T. H. Berlin, and M. Kac, "The spherical model of a ferromagnet", *Phys. Rev.* **86,** 821 (1952).
22. G. S. Joyce, "Critical properties of the spherical model, Vol. 2, Chapter 10, 375, in Phasetransitions and critical phenomena", ed. C. Domb and M. S. Green, *Academic Press* New-York, 1972.
23. R. J. Baxter, "Exactly solved models in Statistical Mechanics", *Academic Press*, London 1989.
24. H. E. Stanley, *"Introduction to Phase Transitions and Critical Phenomena"* (Oxford U. Press, Oxford, UK, 1971).
25. H. E. Stanley, "Exact solution for a linear chain of isotropically interacting classical spins of arbitrary dimensionality", *Phys. Rev.* **179,** 570 (1969).
26. A.E. Siegman, "Lasers", University Science Books, Mill Valley California,1986.